\documentclass[aps,pra,twocolumn,showpacs,superscriptaddress]{revtex4}
\usepackage{bm,amsmath,amssymb,latexsym,graphicx,enumerate}
\usepackage{epsfig}
\usepackage{float}

\newcommand{\be}{\begin{equation}}
\newcommand{\ee}{\end{equation}}
\begin{document}

\title{Radial sine-Gordon kinks as sources of fast breathers }
\author{J.-G. Caputo}
\email{caputo@insa-rouen.fr}
\affiliation{
Department of Mathematics,\\
University of Arizona,\\
Tucson, AZ, 85719, USA}
\author{M.P.  Soerensen}
\email{mpso@dtu.dk}
\affiliation{ Department of Mathematics \\
Technical University of Denmark\\
DK-2800 Kgs. Lyngby, Denmark}

\date{ \today  }

\begin{abstract}
We consider radial sine-Gordon kinks in two, three and higher dimensions.
A full two dimensional simulation showing that azimuthal perturbations
remain small allows to reduce the problem to the one dimensional radial
sine-Gordon equation. We solve this equation on an interval $[r_0,r_1]$
and absorb all outgoing radiation. Before collision the kink is well described
by a simple law derived from the conservation of energy. In two
dimensions for 
$r_0 \le 2$, the collision disintegrates the kink into a fast breather while
for $r_0 \ge 4$ we obtain a kink-breather meta-stable state where breathers
are shed at each kink "return". In three and higher dimensions $d$
a kink-pulson state appears for small $r_0$. The three
states then exist as shown by a study of the $(d,r_0)$ parameter space.
On the application side, the kink disintegration opens 
the way for new types of terahertz microwave generators.

\end{abstract}

\maketitle

PACS : nonlinear dynamics of solitons 05.45.Yv, microwaves 84.40x

\section{Introduction}

The sine-Gordon equation is an important
model both for theory and for applications. It corresponds to 
a classical field with degenerated ground states ($2 n \pi$) \cite{eilbeck}.
In one space dimension, it is integrable
via the inverse-scattering transform and it has two main classes
of solutions, the kink and the breather. The former is particularly
interesting because it is a topological defect separating two
regions where the solution is 0 and $2 \pi$. In higher dimensions
one can introduce the radial kink i.e. a kink which only depends on the 
radius $r$ and this was studied by a number of authors.
Among these Christiansen et al \cite{Christiansen79},
\cite{Lomdahl80}, \cite{Christiansen81} have shown that
such kinks with initial velocity exhibit the return effect where they "grow"
up to some radius and then shrink back. 
Note also the remarkable work by Geicke who described solutions of the
radial sine-Gordon \cite{Geicke82,Geicke84} and indicated that radial kinks 
are destroyed \cite{Geicke83} at the origin in two dimensions. 
This observation was also reported by Bogolubsky and Makhankhov \cite{bm76}.
This particular phenomenon is not well understood. Geicke \cite{Geicke83}
reports in particular a difference in the collapse of the kink in two
and three dimensions.
To analyze these radial
solutions one can assume that the radial term is small so that the
system is a perturbed one dimensional sine-Gordon equation.
One of the analytical tools is a perturbation theory based on inverse 
scattering, see the formulation by Maslov \cite{Maslov85}. When the radius 
is small, however, this
breaks down and only conservation laws can be used so the analysis
becomes very difficult. For example see the techniques used by
Alfimov and Vazquez \cite{Alfimov00} to study the long lived
radial breathers, so-called "pulsons". Using this combination of
analysis and numerics, they showed that 
these waves decay very slowly and in fact do not exist as such\cite{Alfimov00}.

On the application side, the two dimensional sine-Gordon equation 
describes the electrodynamics of a
Josephson junction between two superconducting films in the absence
of external current and dissipation \cite{barone}. The wave part
comes from Maxwell's equations and the sine nonlinearity from
Josephson's constitutive relation.
The variable is the phase difference (or flux) $\psi$ between 
two superconductors. 
In this context the kink solution, a "fluxon" carries a flux quantum
which generates micro-wave radiation in the terahertz range when it collides
on the boundary of the device. 
When the lateral geometry of the device is reduced, the fluxon, once created, is "dragged" towards the narrow edge.
This suggested a design of a particle detector \cite{Nappi95} and also
gave rise to the so-called Eiffel junctions with exponential tapered
width\cite{bcs96,bcs00}. For these, analysis and a preliminary 
experimental realization \cite{Carapella02} confirmed that 
no magnetic field is needed to move the kink, current alone suffices.
The dynamics was shown to be very regular contrary to the standard
rectangular design. Note also the analysis of the resonances
by Jaworski \cite{j05}.

There is a strong link between this Eiffel design 
and the radial sine-Gordon model as we will see below.
This link, together with the formal studies and the applications 
inspired us to 
undertake a numerical study of dynamics of two-dimensional and higher
dimensional radial kinks. Since the theory is difficult we
relied strongly on numerical studies using a careful procedure.
We first solved the two dimensional (2d) sine-Gordon equation for a radial
kink and showed that azimuthal perturbations remain small. This
justifies the reduction to the radial sine-Gordon equation.
We studied this equation numerically for a radial kink initial condition
on a finite domain $r_0 < r < r_1$, and absorbed all 
outgoing radiation. This last point is important because the radiation
reflecting from the boundary and coming back into the computational
domain can perturb strongly the solution.
We varied systematically $r_0$ from 10 to 0 to see
how the radial term $(d-1)\phi_r / r$ in the Laplacian affects the collision.
The dimension $d$ is another parameter that we varied and this shows
new effects. 
We consider the radial term as a perturbation
of the one dimensional sine-Gordon equation and changing $r_0$ we change
the magnitude of this perturbation from small to very large.

Before the collision at $r_0$, we find that the 2d and 3d radial kinks are
well described by a simple equation for the radius obtained from
energy conservation. When collision occurs, the kink is always strongly
affected when $r_0 \le 5$. In two dimensions for $r_0 \le 2$ it 
disintegrates into a fast breather rapidly ejected away from $r_0 $.
For larger $r_0$ we observe a semi-stable kink-breather bound state
which sheds fast breathers at each "return". In all cases the kink
decays to 0. Interestingly in three dimensions for $r_0 = 5$ we
recover the total destruction of the kink. For $r_0 \le 5$ 
all the kink energy cannot be converted into a 
single breather because 
the radial term is too strong to prevent it
from escaping. Instead we observe a kink-pulson bound state that
ejects small high-frequency (low energy) breathers. 
For $d\ge 3$ the collision can yield the three states as shown by
a study of the $(d,r_0)$ parameter plane.
This scenario explains the differences observed by Geike \cite{Geicke83} and
other authors for the two and three dimensional kink collision.
It also opens an avenue for new microwave devices which transform
a fluxon (kink) into a large microwave pulse. \\
The article is organized as such. In section II we illustrate the collapse of a 
sine-Gordon kink in two dimensions and show that there are no azimuthal
effects. This justifies the reduction to the radial sine-Gordon equation.
Section III is the study of its conservation laws. Using these we deduce a 
simple model for the shrinking. We examine in detail the radial kink
collision in two, three and higher dimensions in section IV and characterize
the emission of breathers. We conclude in section V and suggest a design
for a terahertz radiation source.

\section{The collapse of a Sine-Gordon kink in a 2d sector} 

\begin{figure} [H]
\centerline{ \epsfig{file=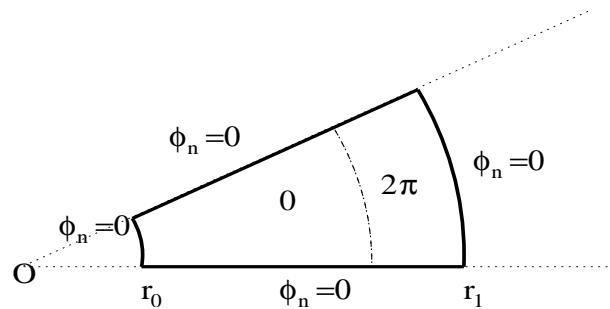,height=4 cm,width=8 cm,angle=0}}
\caption{Sketch of the 2d domain for the sine-Gordon equation. The radial kink
initial condition is shown as a dotted-dashed line.}
\label{sketch}
\end{figure}

To illustrate the problem that we will consider we present here 
a 2d numerical study of the dynamics of a radial kink in a sector.
The 2d sine-Gordon equation reads 
\be\label{2dsg}
\psi_{tt} - {\psi_{\theta \theta} \over r^2} - \psi_{rr} -{\psi_r \over r} + \sin \psi =0.
\ee
As domain we consider the sector $r_{0}<r<r_{1}$, $0<\theta<\theta_0$
shown in Fig. \ref{sketch}. The boundary conditions corresponding to no external current
are homogeneous Neuman so that $\psi_{\theta}=0$ for $\theta=0,\theta_0$
and $\psi_r=0$ for $r=r_{0},r_{1}$. 

We consider the propagation of a sine-Gordon kink inside such a sector.
In that case the initial condition is given by \cite{eilbeck}
\be\label{radkink}
\phi(r,t=0)= 4 {\rm atan}(\frac{r- R_0}{\sqrt{1-u_0^2}}) ,\ee
where $R_0,u_0$ are respectively the initial position and velocity of the
kink.
\begin{figure}[H]
\vskip -1.5 cm
\centerline{ \epsfig{file=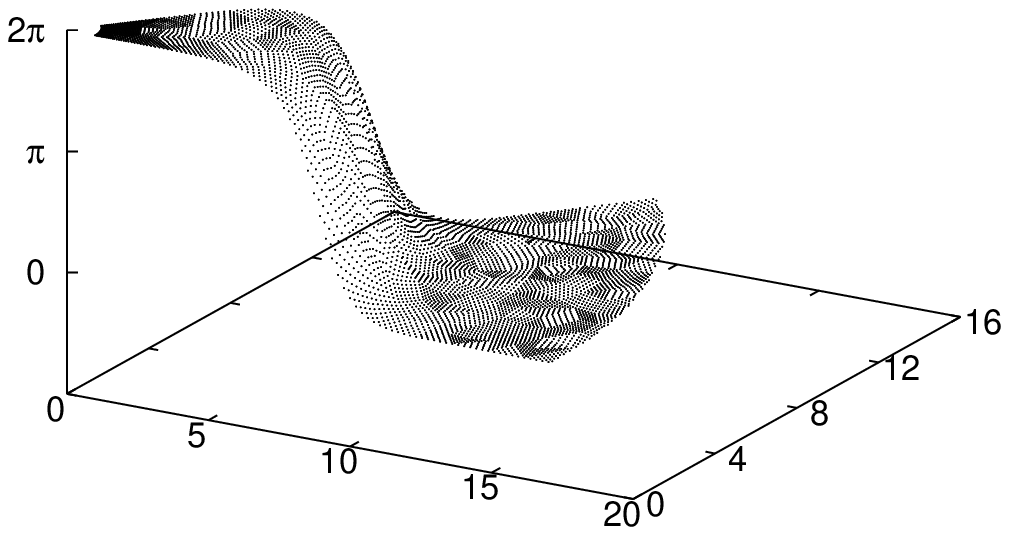,height=5 cm,width=8 cm,angle=0}}
\vskip -1.7 cm
\centerline{\epsfig{file=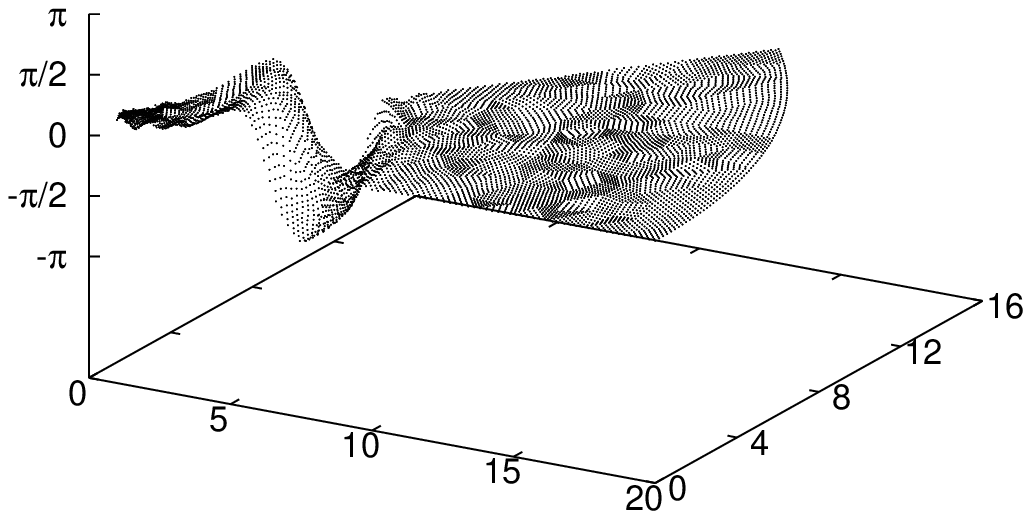,height=5 cm,width=8 cm,angle=0}}
\vskip -.7 cm
\caption{Snapshots of the evolution of a kink for the 2d 
sine-Gordon equation in a 
wedge. The kink is started at $R_0=12$ in a domain such that 
$r_{0}=1,r_{1}=20$. The top panel shows 
the solution for $t=9$ before collision and 
the bottom panel 
the solution for $t=25$ 
after the collision. The $z$ range for the top plot is $[0;2\pi]$
while for the bottom plot it is $[-\pi;\pi]$.}
\label{2d}
\end{figure}

We have computed the evolution of such an initial condition using the
comsol finite element software \cite{comsol}. Fig. \ref{2d} shows two
different snapshots of the evolution
of a kink in a domain such that $r_{0}=1,r_{1}=20$ assuming 
$R_0=10$ and $u_0=0$.
On the top panel, showing the kink at $t=9$, the kink is accelerated 
towards the narrow edge. Notice the absence of radiation and the characteristic
overshoot. 
The bottom panel presents the solution after collision $t=25$ and shows that
the kink has disappeared and only a flat background $\phi \approx 2\pi$
persists with some oscillations. Despite the violence of the collision
all the energy remains in the radial $n=0$ mode and no azimuthal
modes are excited. 
The space discretisation (finite elements) does not preserve the radial
symmetry. Nevertheless azimuthal perturbations do not grow. We therefore
now look for a reduction of the model to the radial case and justify this
approximation.

To reduce the 2d problem it is natural to expand
in azimuthal modes using the cosine Fourier series
\be\label{aziexp}
\psi(r,\theta,t)=\sum_{n=0}^\infty \phi_n(r,t) \chi_n(\theta),\ee
where $\chi_n(\theta)=\cos(\frac{n\pi\theta}{\theta_0})$.
Plugging the expression (\ref{aziexp}) into (\ref{2dsg}) and 
projecting onto $\chi_0$ we obtain the evolution of $\phi_0$
\be\label{evophi0}
-{\phi_0}_{tt} + {\phi_0}_{rr} +\frac{{\phi_0}_r}{r}=\frac{1}{\theta_0}
\int_0^{\theta_0} \sin(\phi_0 + \phi_1 \chi_1 + \phi_2 \chi_2+ \dots ) d\theta.\ee
The integrand in the right hand side can be written as
$$\sin(\phi_0 + \phi_1 \chi_1 + \dots) = 
\sin(\phi_0) \cos (\phi_1 \chi_1 + \phi_2 \chi_2+ \dots )$$
$$+\cos(\phi_0) \sin (\phi_1 \chi_1 + \phi_2 \chi_2+ \dots ) .$$
The integral on the right hand side of (\ref{evophi0}) becomes
$${\sin(\phi_0) \over \theta_0}\int_0^{\theta_0}d\theta \cos (\phi_1 \chi_1 + \phi_2 \chi_2+ \dots ) $$
$$+ {\cos(\phi_0) \over \theta_0} \int_0^{\theta_0}d\theta \sin (\phi_1 \chi_1 + \phi_2 \chi_2+ \dots ) 
  . $$
To estimate these terms we expand the cosine and sine. Then we find that
the non zero contribution for the first term will yield terms
of the form 
$$ {\phi_i^2 \over 2}\sin(\phi_0), $$
and will yield cubic terms for the second integral.
This shows that if the $\phi_i$ are small, one can assume that
$$\sin(\phi_0 + \phi_1 \chi_1+ \dots) \approx \sin(\phi_0 )+\cos(\phi_0) (\phi_1 \chi_1 + \phi_2 \chi_2+ \dots ), $$  
so that (\ref{evophi0}) reduces to the radial 1D sine-Gordon equation
\be\label{radsg}
-{\phi}_{tt} + {\phi}_{rr} +\frac{{\phi}_r}{r}=\sin(\phi),\ee
where the $0$'s have been omitted for simplicity.
The model (\ref{radsg}) can be obtained for any angle $\theta_0$
and in particular for the whole two-dimensional sector. It is also linked to the
variable width sine-Gordon equation which contains
the term $\phi_x ~ w'(x)/w(x)$ \cite{bcs96}. The radial
sine-Gordon corresponds to $w(x)=\theta_0 x$  while the
Eiffel junction is for $w(x)= w_0 e^{-\lambda x}$.

\section{Numerical procedure }

We now detail the numerical procedure because it is the basis of this
work. Another reason is that the approximate analysis based on
perturbation methods is difficult to validate a priori for small
radiuses. We give it meaning by comparing the predictions to the
numerical solution. 
We solve the radial sine-Gordon equation using the
method of lines where the space discretisation is done using 
a finite difference and the time advance is done using an ode solver
(DOPRI5 ordinary differential equation solver \cite{Hairer}). This
method is flexible and one can increase easily the space discretisation.
Another time integrator we have used for comparison is the Verlet
method \cite{lr04}.
The number of discretisation points for a typical run is 4000 and the
accuracy is checked by computing the hamiltonian H (\ref{en_radsg}). For
all cases presented the relative error is smaller than $2 10^{-5}$.
The boundary condition at $r_{0}$ is of the Neuman type so that
there is perfect reflection.
When $r_{0}=0$ care must be taken because the operator $\phi_r / r$
should be regularized because we have an indetermination $0/0$. The way to
do this is to invoke the limit
\be\label{regul}
\frac{\phi_r}{r} = \frac{\phi_r(r,t)-0}{r-0}\to_{r\to 0} \phi_{rr}(r=0,t),\ee
so that $\phi_{rr}+\phi_r/r |_{r=0}=2 \phi_{rr}|_{r=0}$.

\begin{figure}
\centerline{ \epsfig{file=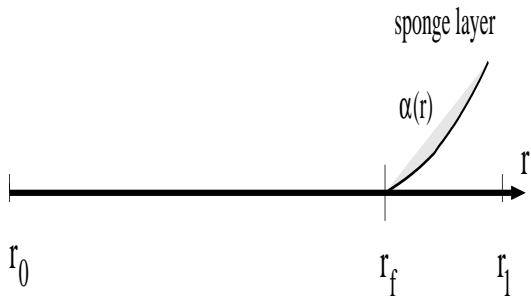,height=4 cm,width=7 cm,angle=0}}
\caption{Sketch of the 1D computational domain.}
\label{domain}
\end{figure}

At the instant of collision radiation is emitted from the kink. To 
avoid it
re-entering the computational domain we introduce a "sponge layer" where
waves are damped so that the equation becomes
\be\label{gen_radsg2}
{\phi}_{tt} - {\phi}_{rr} -(d-1) \frac{{\phi}_r}{r}+\sin(\phi)=-\alpha(r){\phi}_{t},
\ee
where $\alpha(r)$ increases smoothly from $r_d=0.9 r_1$ to the edge 
of the domain $r_1$ as shown in Fig. \ref{domain}. This mechanism
kills all radiation that travels to the right and exits the 
computational domain.
The amount of energy leaving the computational domain
is then computed using the flux relation (\ref{flux_en}) evaluated at
$r_f < r_d$. This "sponge layer" is better adapted for our purposes
than a perfectly patched layer because it damps all out-going waves
not only the ones of speed one.

The equation (\ref{radsg}) is not integrable as in the 1d case and there
are only a finite number of conservation laws. These are the main
analytical tool to study the solution. We consider them in the next section.

\section{Conservation laws}

The radial sine-Gordon equation in $d$ dimensions
\be\label{gen_radsg}
{\phi}_{tt} - {\phi}_{rr} -(d-1) \frac{{\phi}_r}{r}+\sin(\phi)=0,\ee
in a finite domain $[r_{0},r_{1}]$
possesses the following energy conservation law.
\be\label{en_cons}
{dH \over dt}  = [r^{d-1} \phi_r \phi_t]_{r_{0}}^{r_{1}},\ee
where the Hamiltonian $H$ is
\be\label{en_radsg}
H = \int_{r_{0}}^{r_{1}} r^{d-1} dr 
[ {\phi_t^2 \over 2} + {\phi_r^2 \over 2} + (1-\cos \phi )]\equiv \int_{r_{0}}^{r_{1}} r^{d-1} dr {\cal H}. \ee
To see this multiply (\ref{gen_radsg}) by $\phi_t$, integrate
over the domain and integrate by parts the $\phi_r$ term. 
Assuming a Neuman boundary condition at $r=r_{0}$ we naturally
obtain the flux relation for the energy
\be\label{flux_en1}
{dH \over dt}  = {r_{1}}^{d-1} \phi_r \phi_t|_{r_{1}} .\ee
By integrating this relation over time we obtain
\be\label{flux_en}
H(t) = H(0)+ \int_0^t \phi_r(r_1,t') \phi_t(r_1,t')  dt' \ee
This enables us to compute how much energy leaves the
computational domain at $r=r_f$. The energy conservation will
be crucial to explain many properties of the solution.

Another conservation law is related to the momentum $\Pi$ of the
wave defined as
\be\label{momentum}
\Pi = \int_{r_{0}}^{r_{1}} r^{d-1} dr
\phi_t.\ee
From the partial differential equation (\ref{gen_radsg}) we get 
 \be\label{flux_momentum}
{d\Pi \over dt}  = [r^{d-1} \phi_r ]_{r_{0}}^{r_{1}} 
-\int_{r_{0}}^{r_{1}} r^{d-1} dr \sin \phi ,\ee
which shows that even for Neuman boundary conditions at
$r=r_{0},r_{1}$ the momentum is not conserved.

For a localized wave such as a kink, the integrands in $H$ and $\Pi$
are highly localized in $r$.
In 2d we can get a good approximation of the motion before collision
by using the kink solution to the 1D sine-Gordon equation as an ansatz
\be\label{sgk}
\phi_k = 4 {\rm atan} [\exp({r-R \over \sqrt(1-{\dot R}^2)})], \ee
where $R$ is the kink position. 
We assume that the kink is not too close to the boundary so the
integral can be taken from $-\infty $ to $\infty$. 
\begin{figure}
\centerline{ \epsfig{file=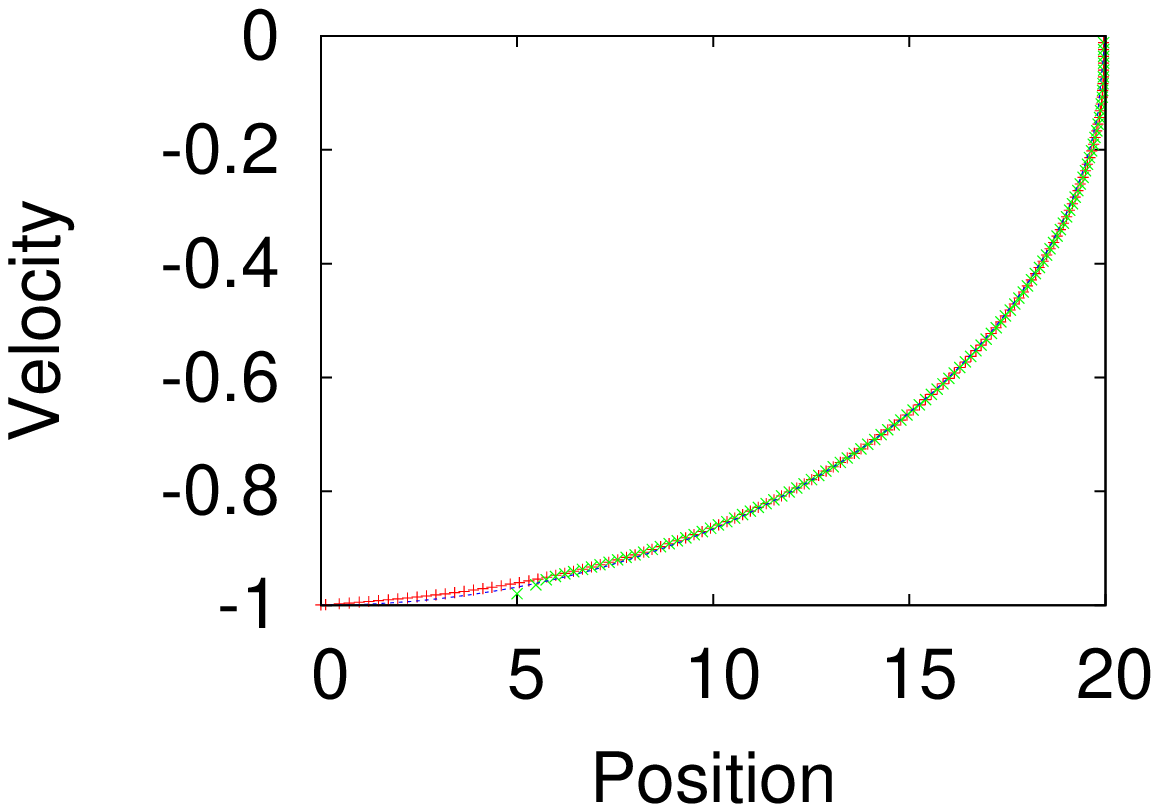,height=4 cm,width=9 cm,angle=0}}
\centerline{ \epsfig{file=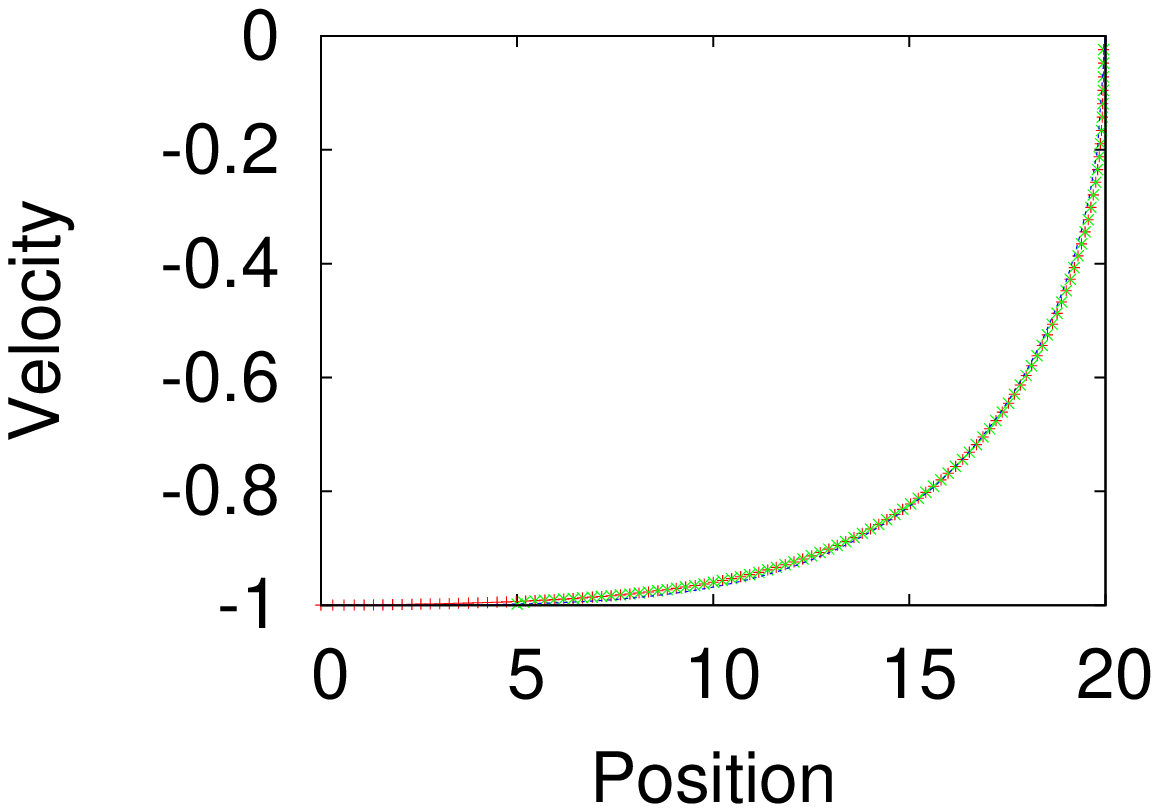,height=4 cm,width=9 cm,angle=0}}
\caption{Top panel: 2d radial kink velocity for $r_{0}=0,5$ 
as a function of the position $R$.
The velocity obtained from
the numerical solution is given by the points and the analytical
estimate (\ref{v2}) is given by the dashed line. Bottom panel: 3d radial
kink velocity for $r_{0}=0,5$ as a function of the position $R$. 
The velocity obtained from
the numerical solution is given with the points and the analytical
estimate (\ref{v3}) is given by the dashed line. }
\label{speed}
\end{figure}
To calculate it we change the integration variable so that
$r-R = r'$ and write
$$H = R \int_{-\infty}^{+\infty} {\cal H} dr'  
+ \int_{-\infty}^{+\infty} r' dr'{\cal H}. $$
The second integral is then 0 because of parity.
This gives 
\be\label{sghama}
H \approx 8 {R \over \sqrt{1-({\dot R})^2}}.\ee
At $t=0$ we start the kink at $R=R_0$ with $u_0$ initial velocity so
that $H= 8 R_0$. From that one can compute the velocity $u={\dot R}$
as a function of $R$
\be\label{v2}
u = \pm \sqrt{1-({R \over R_0})^2 (1-u_0^2)}.\ee
Fig. \ref{speed} shows the velocity $u$ vs the position $R$ of a 2d radial
kink (top panel) and the 3d radial kink (bottom panel)
obtained from the numerical solution 
as it propagates towards $r_0$ for $r_0=0,1,5$ and 10. 
The velocity is estimated by assuming the kink functional 
dependance (\ref{sgk}). One then estimates 
$${\rm Max} (\phi_t) = {2u  \over {1-u^2}},$$
and deduces the velocity $u$.
On the same plot we report the estimate (\ref{v2}). The picture shows that the
agreement is very good even when the kink is close to $r_0$.
Assuming the wave keeps it's kink profile, its position $R$ follows
the differential equation obtained from (\ref{sghama})
\be\label{kink_ode2}
{\ddot R} + {R \over R_0^2} =0.\ee
whose solution is 
\be\label{skink_ode2}
R(t) = R_0 \cos {t \over R_0^2} . \ee
As expected \cite{bcs96} the kink travels towards the "narrow" end 
and accelerates.

Using similar arguments we can reduce the Hamiltonian for the
3d kink to
$$H = R^2 \int_{-\infty}^{+\infty} {\cal H} dr'  
+ \int_{-\infty}^{+\infty} r'^2 dr'{\cal H}. $$
The first term gives 
\be\label{sghama2}
H \approx 8 {R^2 \over \sqrt{1-({\dot R})^2}}.\ee
The second term is a small correction of the order of the cube
of the "width" of the kink. It is much smaller than the
leading term (\ref{sghama2}).
We then obtain the evolution of the 3d radial kink as
\be\label{v3}
u = \pm \sqrt{1-({R \over R_0})^4 (1-u_0^2)}.\ee
The comparison of (\ref{v3}) with the numerical solution is also
very good as shown in the bottom panel of Fig. \ref{speed}.
The differential equation for $R$ is 
\be\label{kink_ode3}
{\ddot R} + {2 R^3 \over R_0^4} =0,\ee
which implies a decay of $R$ but does not have a simple solution.

\section{Radial kink collision in 2d and 3d}

To understand the collision of a kink with the boundary at $r=r_{0}$
we have conducted extensive numerical studies varying systematically
$r_{0}$. The main result for both 2d, 3d and larger d is that the kink 
does not survive collision in a proper way when $r_0 $ is small. 
It decays after a few collisions and at each collision it emits part of its 
energy in the form of fast breathers that
escape the radial potential. This is true for all initial conditions
$R_0 > 10$. 
This is interesting since it is
a way to destroy a topological defect. The specifics vary from the
2d case to the 3d case so we will consider them separately.
\begin{figure}
\centerline{ \epsfig{file=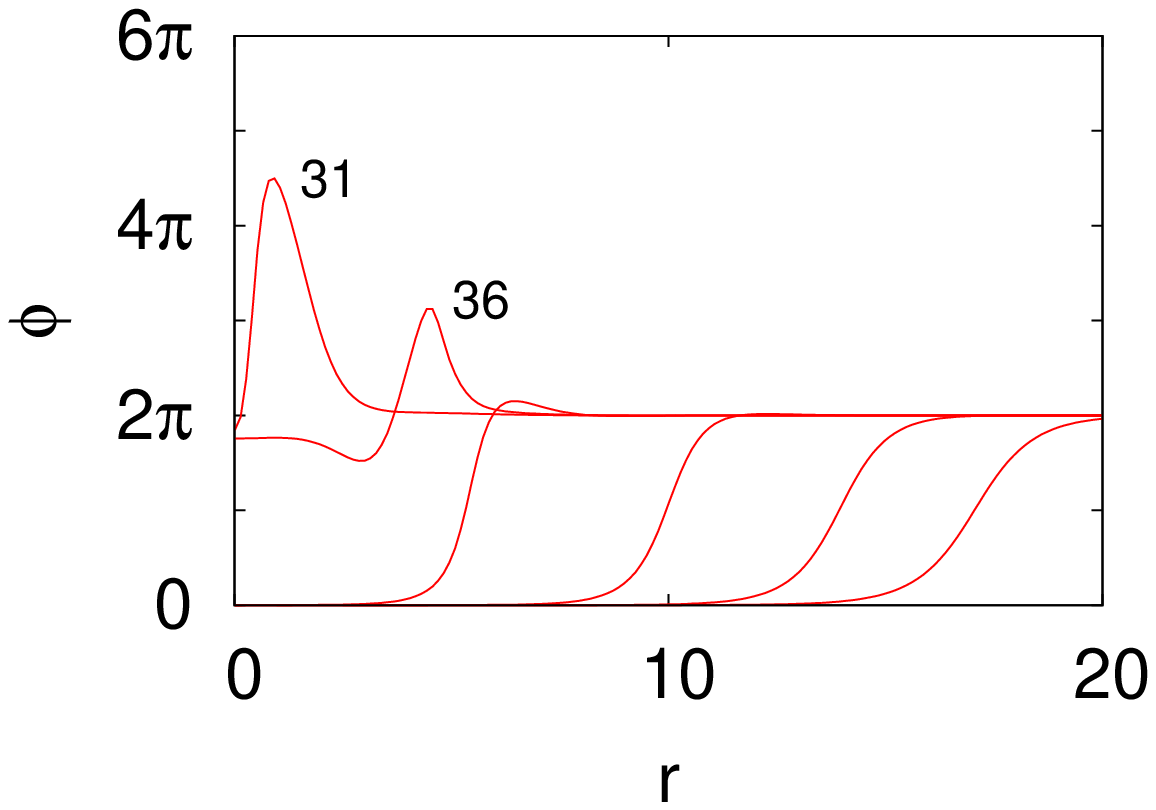,height=5 cm,width=9 cm,angle=0}}
\centerline{ \epsfig{file=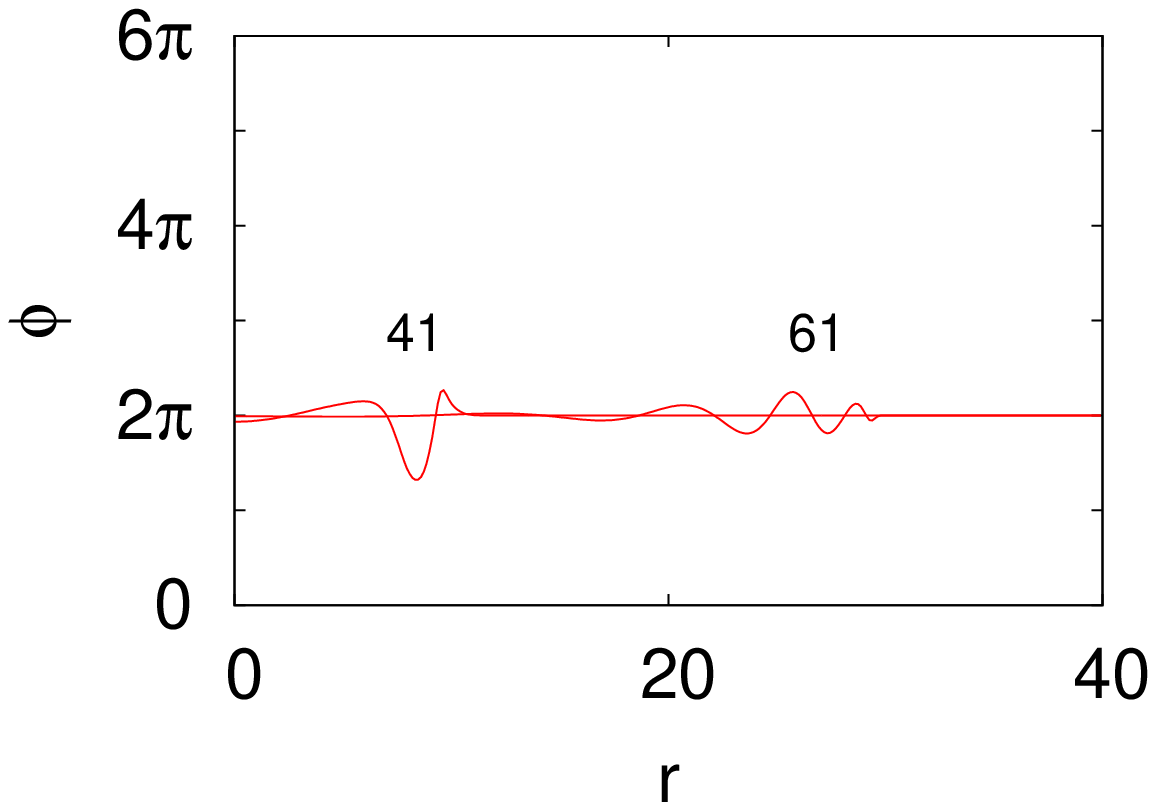,height=5 cm,width=9 cm,angle=0}}
\caption{2d radial kink collision for $r_{0}=0$. Plot of $\phi(x,t)$ as a 
function of $x$ for 
$t=11,~16,~21,~26,~31$ and $36$ (top panel) and $t=41,61$ (bottom panel). 
The kink
is started at $R_0=20$. The parameters are $r_{0}=0,r_{1}=40$.}
\label{2d_uuu_r0}
\end{figure}

\subsection{collision in 2d}

\begin{figure}
\centerline{ \epsfig{file=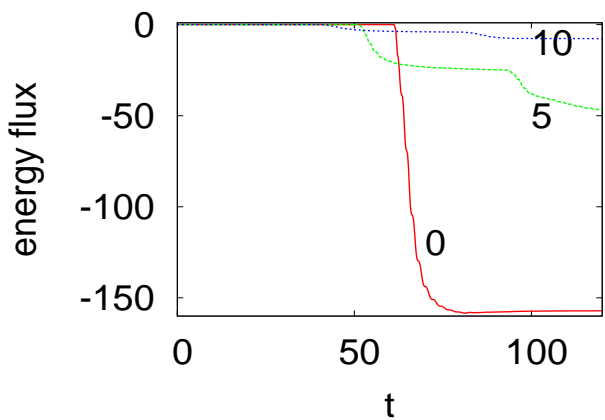,height=6 cm,width=9 cm,angle=0}}
\caption{2d radial kink collision. Time evolution of the energy exiting the
domain at $r_f= 30$, for $r_{0}=0,5$ and 10. The kink is started 
at the same position $R_0=20$
so the initial energy $H_0\approx 160$ is the same for the three cases.}
\label{2d_flux}
\end{figure}
Fig. \ref{2d_uuu_r0} shows snapshots of $\phi(x,t)$ as a function of $x$
for different times before and after the collision at $r_0=0$. The
top panel shows the times $t=11,\dots 36$ corresponding to the kink
being accelerated towards $r_0=0$. Notice the large overshoot for
$t=36$. The right panel of Fig. \ref{2d_uuu_r0} shows the two instants
$t=41$ and $t=61$ showing that very little is left of the initial
kink. There is just a small disturbance around $2\pi$ traveling
towards large $r$. In fact all the kink energy from (\ref{sghama}) 
$H\approx 20\times 8=160$ leaves the computational
domain as shown by the energy flux (\ref{flux_en}) measured at $r=30$ 
shown in Fig. \ref{2d_flux}.

\begin{figure}
\centerline{\epsfig{file=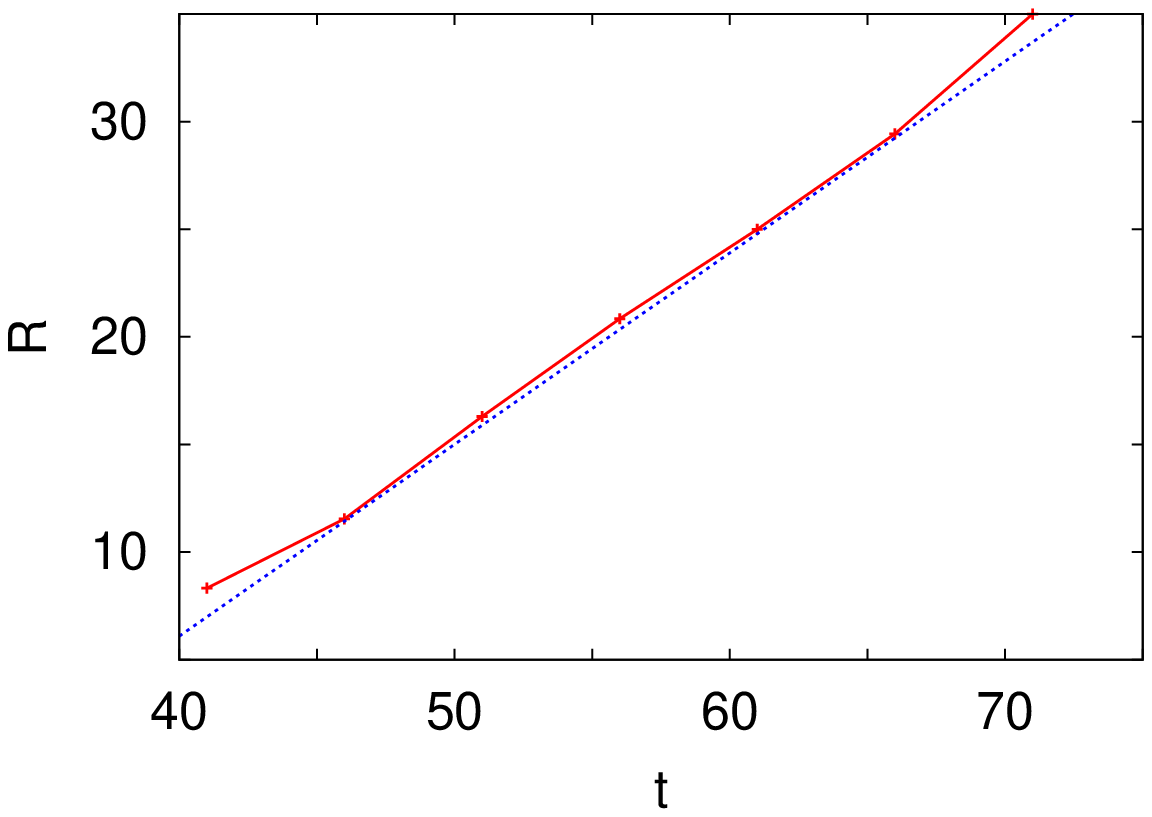,height=5 cm,width=9 cm,angle=0}}
\centerline{\epsfig{file=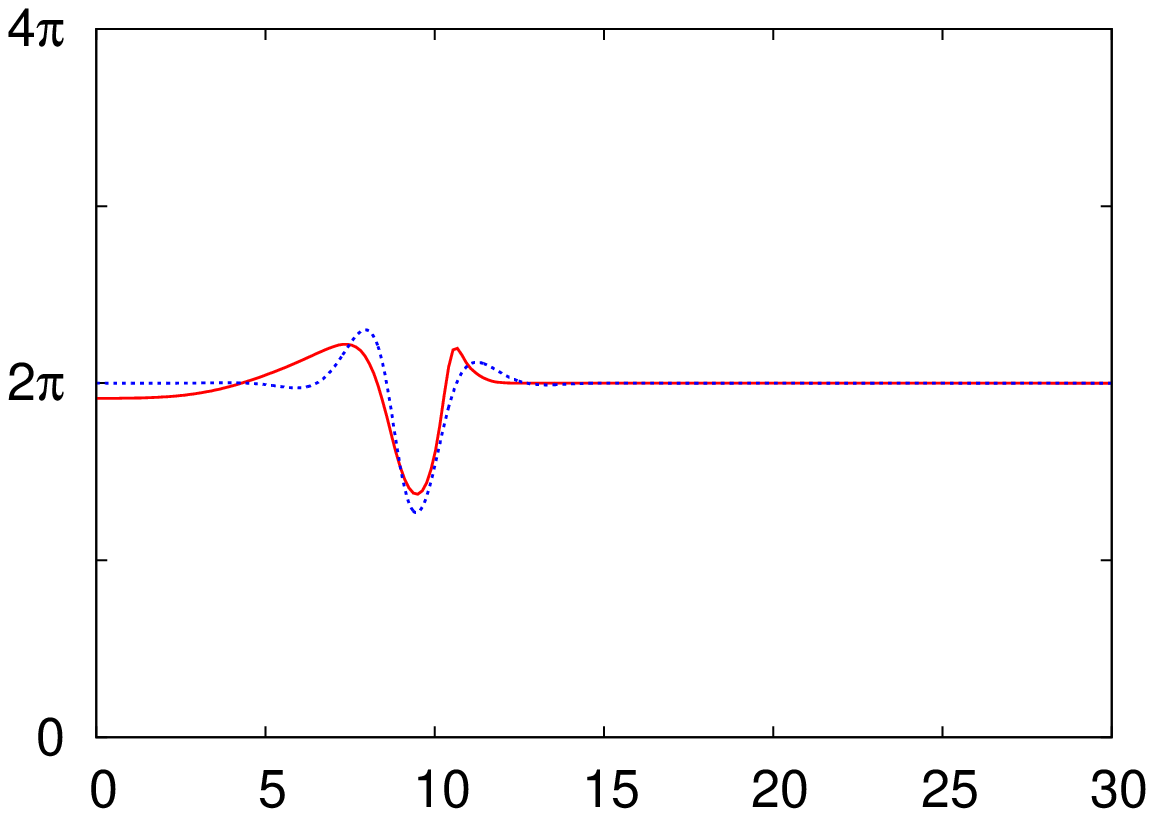,height=5 cm,width=9 cm,angle=0}}
\centerline{\epsfig{file=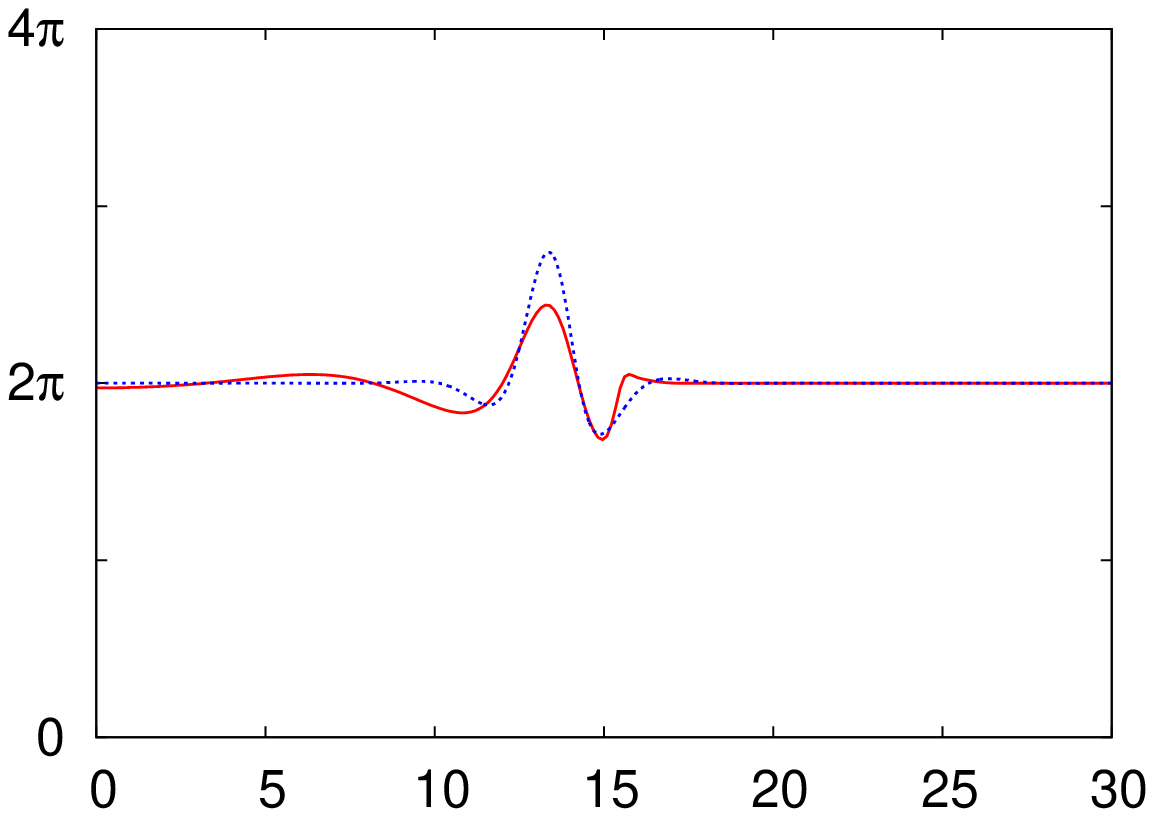,height=5 cm,width=9 cm,angle=0} }
\caption{Characterization of the wave after 2d radial kink collision. The
top plot shows the center of mass of the wave as a function of time.
The middle and bottom plots show the numerical solution together with
the breather fit (\ref{brea}) for $t=41$ and $t=46$ respectively.}
\label{2d_brea_r0}
\end{figure}

The wave present in the domain after the collision is a fast breather. 
Recall that a sine-Gordon breather is given by
\be\label{brea}
\phi_b = 4 {\rm atan} \left [  {\sqrt {1-\omega^2} \cos(\omega \gamma (t-u r -t_0)) \over 
\omega \cosh(\sqrt {1-\omega^2} \gamma (r -R_0-u t))} \right ], \ee
where $\gamma$ is the usual Lorentz factor
$$\gamma = {1 \over \sqrt{1-u^2}} . $$
The energy of the breather on the infinite line is given by \cite{scott}
\[ H = 16 \gamma \sqrt{1-\omega^2},\]
so that using the same argument as for the kink we get for the radial case
\be\label{en_breather_radial}
H_b = 16 R \gamma \sqrt{1-\omega^2},\ee
where $R$ is the center of mass of the breather.
To identify this fast breather in the numerical solution we 
have plotted the position of its
center of mass as a function of time on the top panel of Fig. \ref{2d_brea_r0}.
The velocity estimated by the fit (dashed line) is $u\approx 0.89$. On the middle
and bottom panels we have plotted the analytical expression of the breather 
(\ref{brea})
added to the $2 \pi$ background together with the numerical solution for $t=41$ and $t=46$. 
The parameters used for the fitting are $$\omega =0.82,~~t_0=-3, ~~x_0=9.2$$
As can be seen the fit is very good, the error in the energy between the
fit and the numerical solution is 10 \%.
Such a breather can "escape" the radial trap because its high frequency 
averages out the radial force. Also note that the radial wave equation does
not support any traveling wave like in the 1D case, so any emitted
radiation has to be in the form of a wave packet. It is interesting to
find breather solutions as a product of the disintegration of a kink. 
In the 1d case for which the sine-Gordon equation is integrable, the
breather and kink-antikink pairs are separated. Here, with the radial
term a connection has been opened between the two states. This is
similar to the numerical experiments of \cite{dsv08} where kink-antikink
pairs are created out of a train of small breathers in the $\phi^4$ model
which is a perturbation of the sine-Gordon equation.

\begin{figure}
\centerline{ \epsfig{file=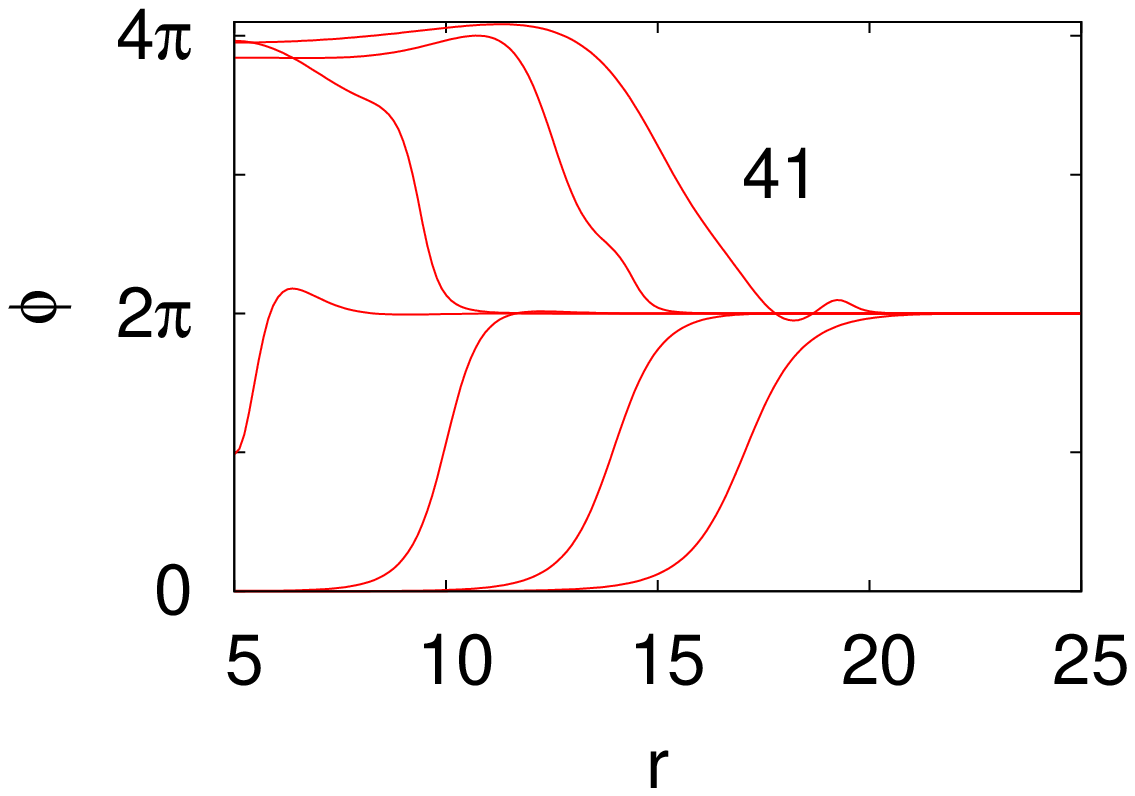,height=5 cm,width=9 cm,angle=0}}
\centerline{ \epsfig{file=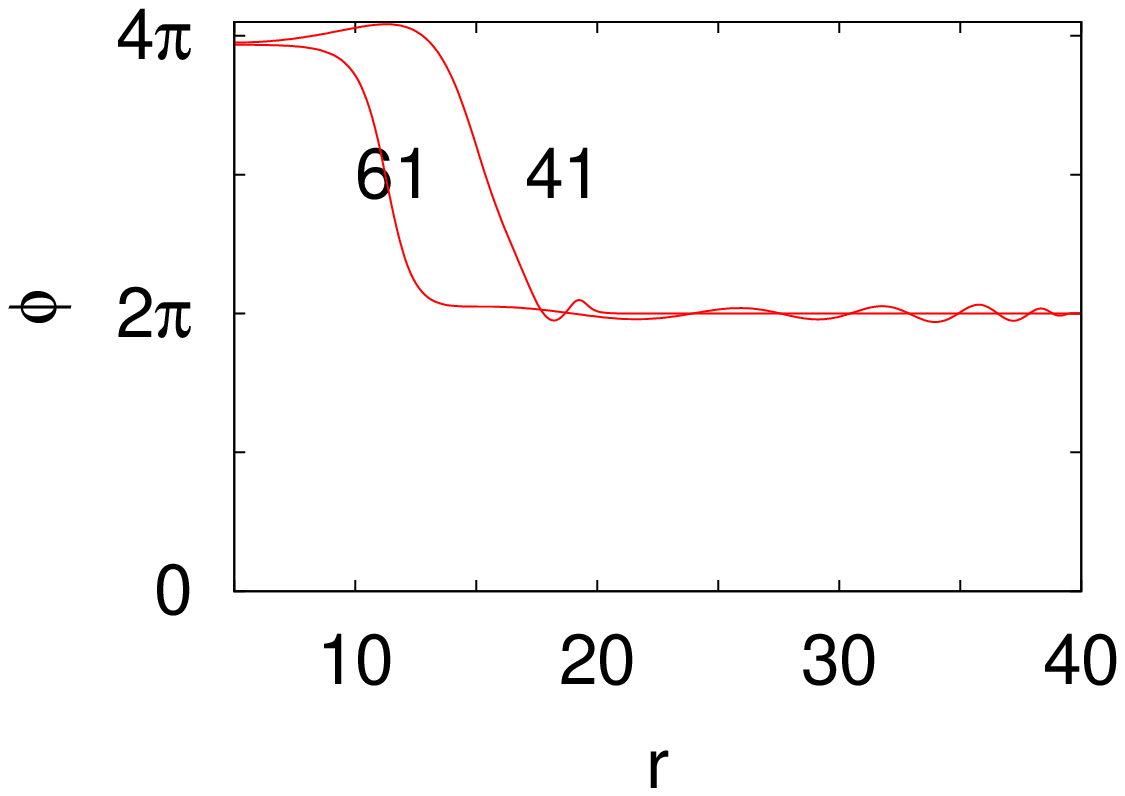,height=5 cm,width=9 cm,angle=0}}
\caption{2d radial kink collision for $r_{0}=5$. The top panel shows
plots of $\phi(x,t)$ as a
function of $x$ for
$t=11,~16,~21,~26,~31,~36$ and $41$. The bottom panel shows 
snapshots of $\phi(x,t)$ for $t=41$ and $61$. The kink
is started at $R_0=20$. The parameters are $r_{0}=5,r_{1}=40$.}
\label{2d_uuu_r5}
\end{figure}

For a larger value of the boundary $r_0=5$, the kink is reflected 
and looks roughly
like an antikink. The snapshots are shown in Fig. \ref{2d_uuu_r5}. 
Notice the
return occurring for $t\approx 41$. There is however about 20 \% of
energy (about 21) lost after the collision,
as shown in the flux plot Fig. \ref{2d_flux}. The approximate antikink that is
formed has an energy which is about $16\times 8= 128$ so that it "stops" at
$R\approx 16$ as shown in the bottom panel of Fig. \ref{2d_uuu_r5}.
Again a breather is emitted, it is bound to the antikink up to $t=41$ after
which it detaches and propagates to the right. 
To characterize this breather
we ran the simulation over a much larger space interval $r_1=100$. 
The  parameters of the breather can be calculated as follows
using the energy loss 
$$H(t=0)-H(t=81)\approx 
 21 =  x_0 16 {\sqrt{1-\omega^2} \over \sqrt{1-u^2}} $$
with $x_0=30$ corresponding to an instant of observation $t=80$.
This gives the following parameters
$$u=0.7,~~\omega=0.9995$$
which are consistent with the energy loss (21) and the duration of the 
breather passing through the boundary
$$\delta t = {\rm width }/ {\rm speed  }= {\sqrt{1-\omega^2}\over \gamma u} 
\approx 32 .  $$
After the emission of this breather, the kink continues to oscillate and decays 
slowly emitting waves at each collision with $r=r_0$.
For larger $r_0$ as shown in Fig. \ref{2d_flux} for $r_0=10$ the kink
decays much slower. For $t=100$ its energy has diminished by
about 10 \%. At every collision some energy is expelled. 
For such value $r_0$ the radial term $\phi_r / r$ is small
and we are close to the one dimensional situation.

\begin{figure}
\centerline{ \epsfig{file=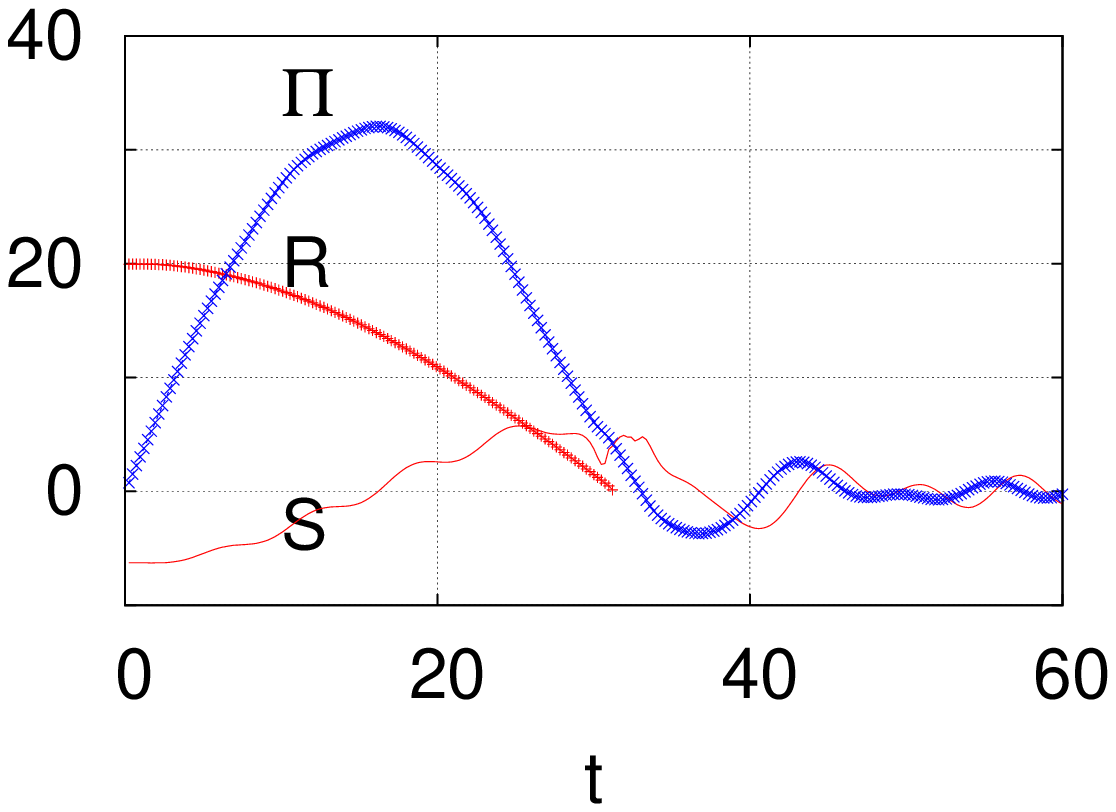,height=3 cm,width=9 cm,angle=0}}
\centerline{ \epsfig{file=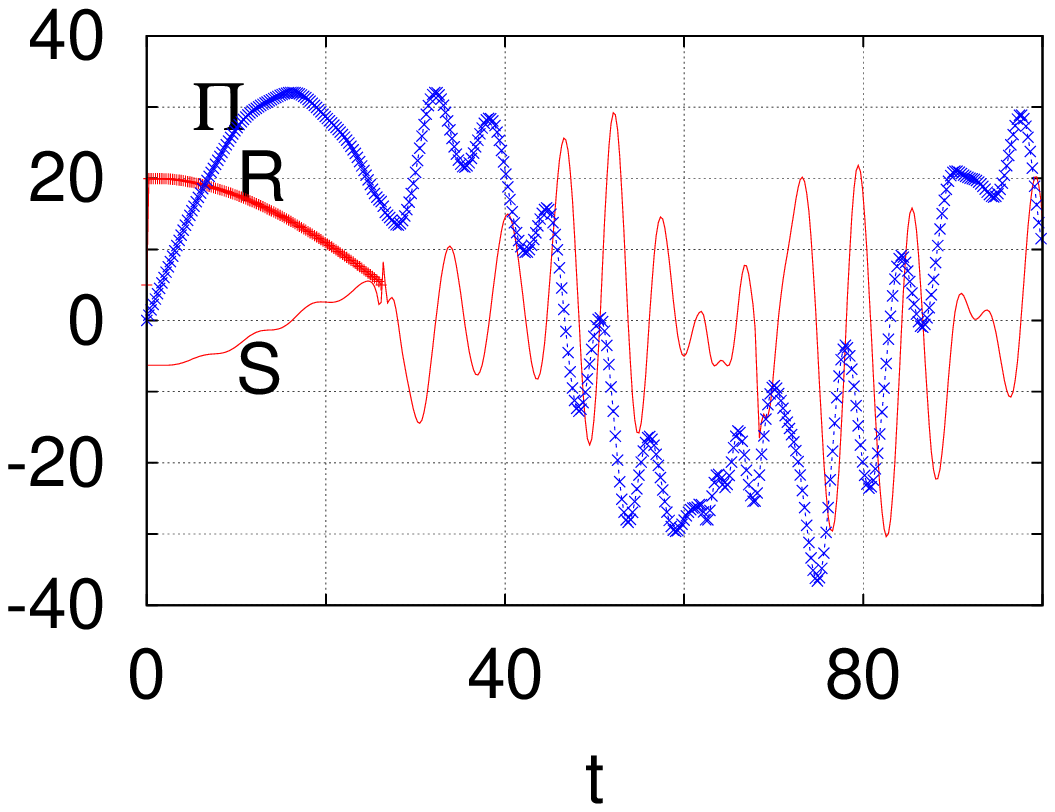,height=3 cm,width=9 cm,angle=0}}
\caption{ 2d radial kink collision. Time evolution of the 
rescaled momentum $\Pi/2$, the kink position $R$,
and the source term $S$ of the momentum equation for 
$r_{0}=0$ (top panel) and $r_{0}=5$ (bottom panel). }
\label{2d_momentum}
\end{figure}

To shed another light on the problem
it is useful to examine the different conservation laws before
and after collision.
We analyzed the difference between $r_0=0$ and $r_0=5$, by computing the
momentum $\Pi$, the right hand side $S$ in the flux of the
momentum  (\ref{flux_momentum}) and the front position $R$. The
latter is defined as the maximum of $\phi_x$. The time evolution
of the three quantities $\Pi,S$ and $R$ are shown in Fig. \ref{2d_momentum}. 
The top panel corresponds to $r_0=0$ and the bottom panel to $r_0=5$.
For $r_0=0$ the momentum $\Pi$ goes through 0 for $R=0$
(collision instant). There $S>0$ so that $\Pi$ will keep decreasing
and oscillate around zero, indicating that the soliton is destroyed. For
$r_0=5$ shown in the bottom panel, $\Pi>0$ at collision and
$S>0$ so that $\Pi$ remains positive for $t\ge 20$. After that instant
$S$ starts to oscillate with a period about 6. $S$ is largely positive
so that $\Pi <0$ on average. We then reach $\Pi=0$ so that the kink
stops at $t\approx 41$ for  which $R=15$. This is the return effect.
\begin{figure}
\centerline{ \epsfig{file=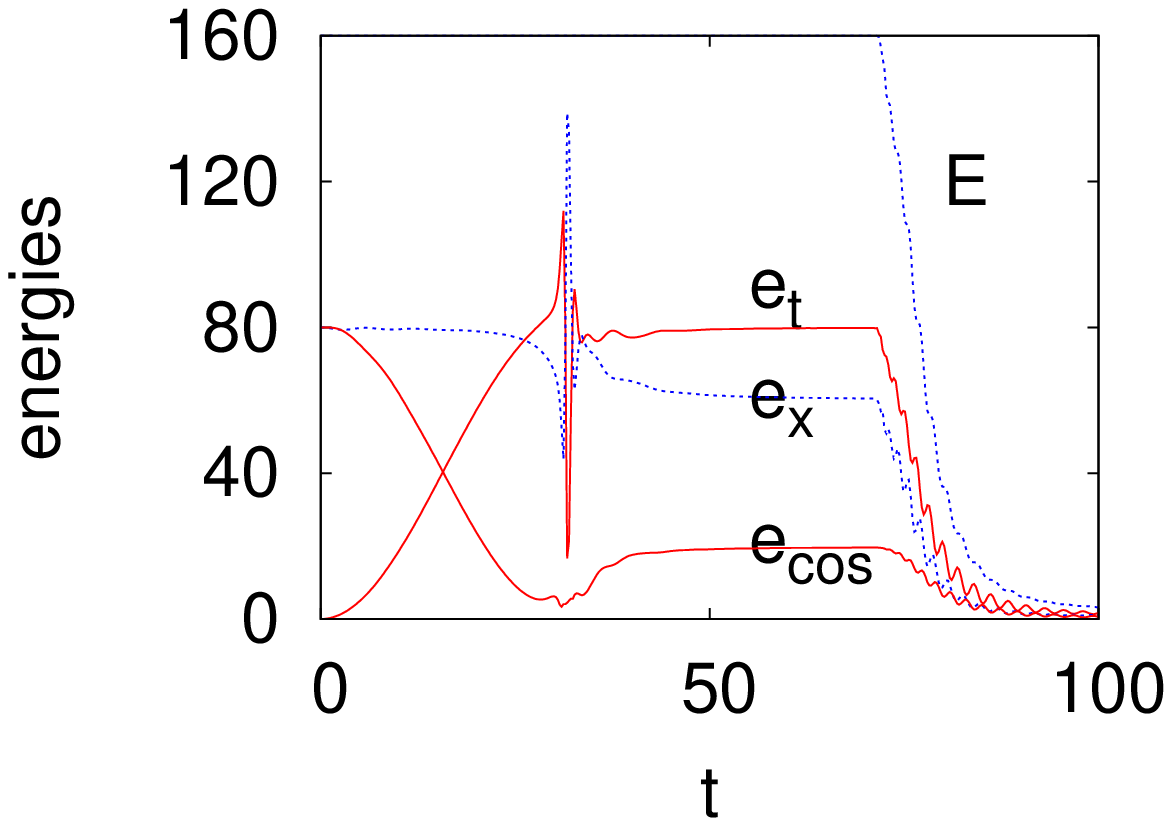,height=5 cm,width=9 cm,angle=0}}
\centerline{ \epsfig{file=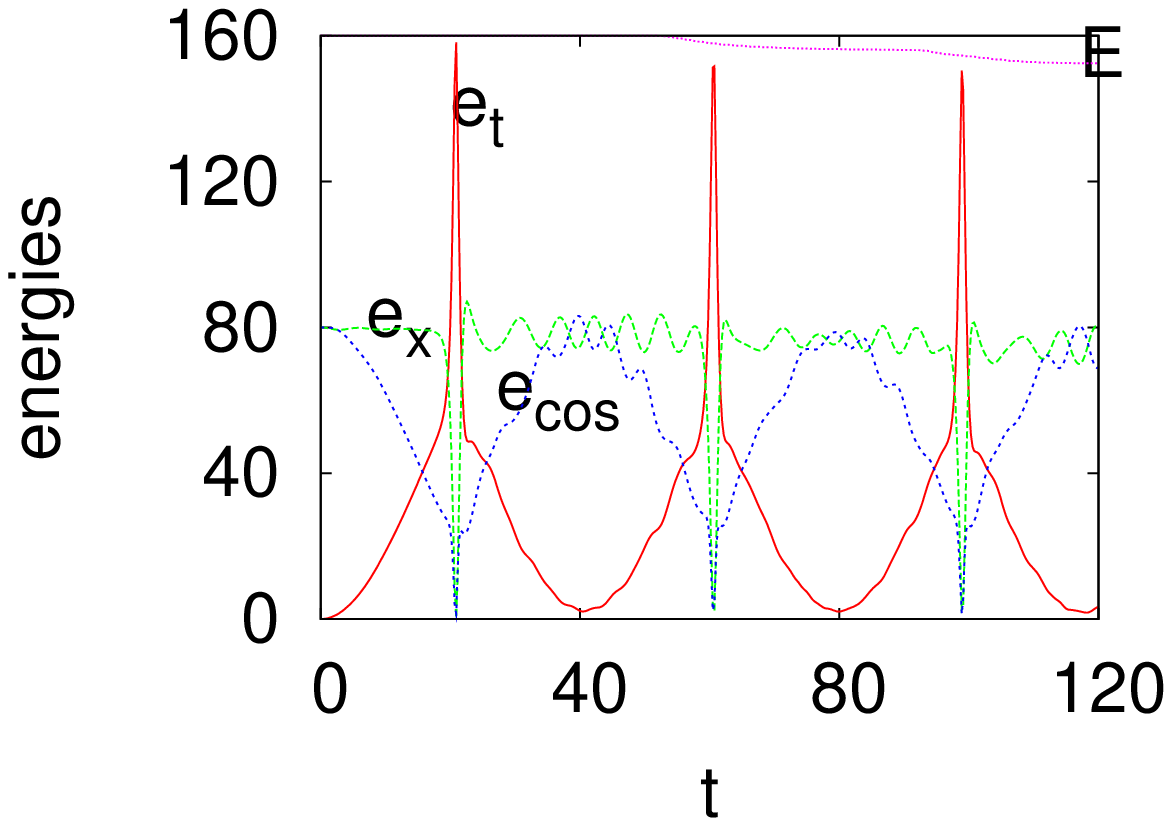,height=5 cm,width=9 cm,angle=0}}
\caption{2d radial kink collision:  Time evolution of the energy components 
$e_t = \int_{r_{0}}^{r_{1}} 0.5 x \phi_t^2 dx $,
$e_x = \int_{r_{0}}^{r_{1}} 0.5 x \phi_x^2 dx $ and $e_p = \int_{r_{0}}^{r_{1}} (1-cos\phi)xdx $ for a kink. On the
top panel $r_{0}=0$ while on the bottom panel 
panel $r_{0}=10$. The kink is started at the same position $R_0=20$.
}
\label{2d_energy}
\end{figure}
The total energy is conserved but gets distributed
differently between the different components,
the kinetic term $e_t = \int_{r_{0}}^{r_{1}} 0.5 \phi_t^2 x~ dx$, the 
gradient term 
$e_x = \int_{r_{0}}^{r_{1}} 0.5 x \phi_x^2 dx $ and the potential 
term $e_p = \int_{r_{0}}^{r_{1}} (1-cos\phi)x~dx$.
Initially the kink has 0 velocity so that $e_t=0$, all the energy
is concentrated in $e_x$ and $e_p$.
For $r_{0}=0$ on the top panel of Fig. \ref{2d_energy}, the kinetic term
$e_t$ increases from 0 to its maximum at the collision and then
remains about constant. The potential term $e_p$ decreases from
its maximum value at $t=0$ and stabilizes around half its value.
The behavior is different for $r_{0}=10$ shown on the bottom panel of
Fig. \ref{2d_energy}, for which the collision is almost elastic. There at the
instant of collision, the
kinetic energy reaches its maximum, the total energy and the potential
energy and gradient energies are almost zero. After collision both
recover their initial values.

\subsection{Collision in 3d and higher d}

\begin{figure}
\centerline{ \epsfig{file=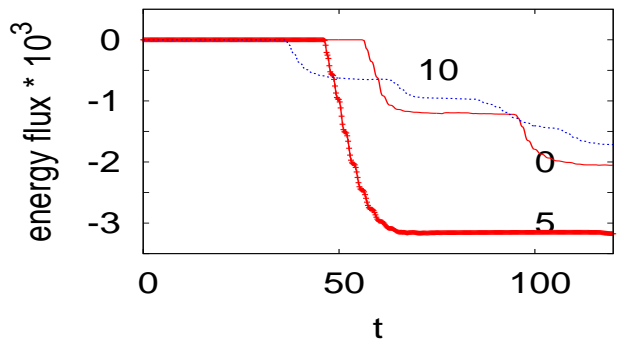,height=5 cm,width=9 cm,angle=0}}
\caption{ Time evolution of the energy flux for three
different values $r_{0}=0,5$ and 10. The kink is started at 
the same position $R_0=20$ so the initial energy $H_0\approx 3200$ 
is the same for the three cases.}
\label{3d_flux}
\end{figure}

\begin{figure}
\centerline{ \epsfig{file=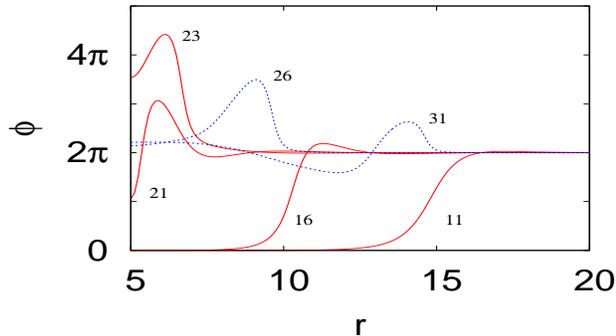,height=5 cm,width=9 cm,angle=0}}
\caption{
3d radial kink collision for $r_{0}=5$:  plot of $\phi(x,t)$ as a
function of $x$ for
$t=11,~16,~21,~23,~26$ and 31. The snapshots after the collision are indicated in
dashed line (blue online).
The kink is started at $R_0=20$. The parameters are $r_{0}=5,r_{1}=40$.}
\label{3d_uuu_r5}
\end{figure}

\begin{figure}
\centerline{ \epsfig{file=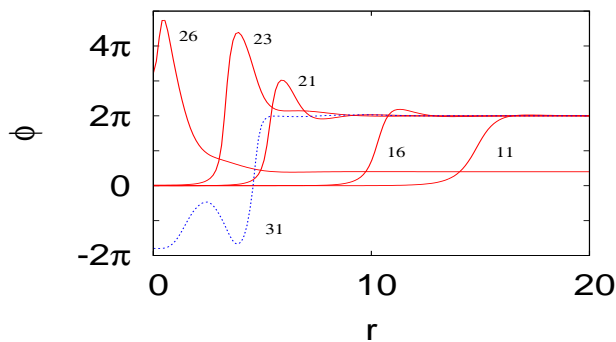,height=5 cm,width=9 cm,angle=0}}
\caption{
3d radial kink collision for $r_{0}=0$:  plot of $\phi(x,t)$ as a
function of $x$ for
$t=11,~16,~21,~23,~26$ and 31. The field for the snapshot $t=23$ has been divided 
by 5 to fit in the plot. The kink
is started at $R_0=20$. The parameters are $r_{0}=0,r_{1}=40$.}
\label{3d_uuu_r0}
\end{figure}

We now consider the collision of a kink in 3d and higher d
to see if there are particular situations. The general result that
radiation is emitted as fast breathers out of the computational domain still
holds. As an example consider the flux of the energy shown for the
3d case in Fig. \ref{3d_flux} for $r_0=0,~5$ and $10$.
Interestingly the case $r_0=5$ is similar to the 2d case for $r_0=0$.
The kink is entirely transformed into a fast breather that exits
the computational domain. This is shown in the series of snapshots
in Fig. \ref{3d_uuu_r5}. The fast breather is clearly seen traveling to the
right at time $t=31$. For $r_0=0$ there is a bound state kink/breather
so that the kink "sheds" a breather at every collision with the boundary and
decays. Fig. \ref{3d_uuu_r0} shows the successive snapshots of the solution
in this case. The solution for $t=31$ is clearly a combination of a kink
and a breather. 
For such a large value of $d$ and such a small $r_0$ the radial term 
$(d-1) \phi_r / r$ is very strong and prevents a low frequency breather from
escaping. Only breathers of frequency $\omega \approx 1$ can escape
and these have fairly low energy. Such a breather will be "shed" from the
kink as it reaches its return point, around $R=5$. The energy and momentum
behave in a very similar way as in the 2d case so we do not present them.

\begin{figure}
\centerline{ \epsfig{file=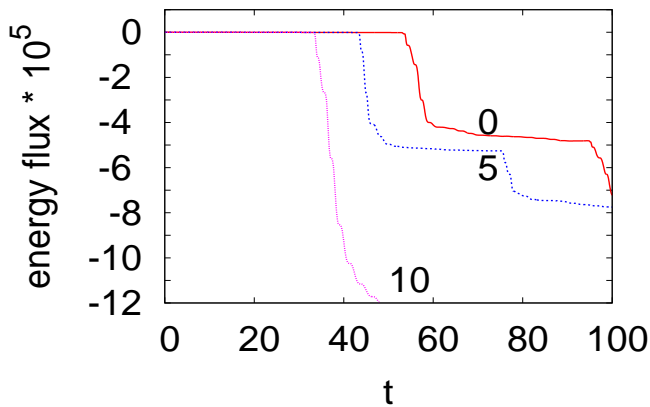,height=6 cm,width=9 cm,angle=0}}
\caption{Energy exiting the computational domain $[r_0: 100]$ for $d=5$
and $r_0=0$ in continuous line (red online) and $r_0=5$ in dashed line
(blue online). The total energy is 1.295743 $\times 10^6$.  }
\label{5d_flux}
\end{figure}

\begin{figure}
\centerline{ \epsfig{file=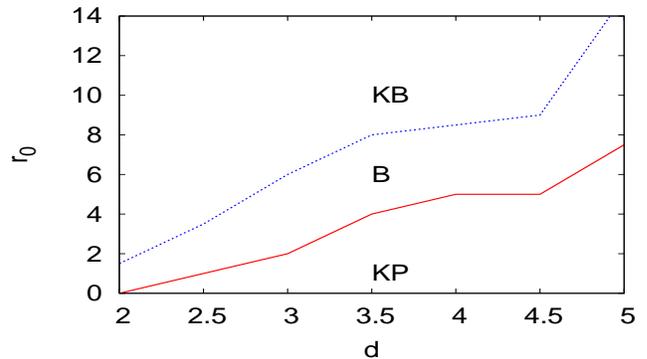,height=5 cm,width=9 cm,angle=0}}
\caption{
Parameter plane $(d,r_0)$ showing the different possible outcomes
for a kink collision, a fast breather (B), a kink-pulson metastable
state (KP) and a kink-breather metastable state (KB).}
\label{dr0}
\end{figure}

To confirm these findings we conducted two simulations with $d=5$,
for $r_0=0$ and $5$ and $r_1=40$. The flux of energy exiting the domain
is shown in Fig. \ref{5d_flux}. It shows two breathers being
emitted respectively at $t=55$ and $t=100$ for $r_0=0$. 
The $(d,r_0)$ parameter plane is shown in Fig. \ref{dr0}. It shows
the coexistence of the three states, the fast breather (B), the
kink-breather metastable (KB)
and the kink-pulson metastable state (KP) 
Interestingly the last one cannot be seen for $d=2$. 

\section{Conclusion}

Motivated by theory and applications, we studied radial sine-Gordon 
kinks in two and higher dimensions.
A full two dimensional simulation showed that azimuthal perturbations
remain small. We therefore reduced the problem to the one dimensional radial
sine-Gordon equation which we solve on an interval $[r_0,r_1]$.
Before collision the kink is well described
by a simple law derived from the conservation of energy. 
In two dimensions the collision of the kink
with the boundary $r_0$ will result in a fast breather
for small $r_0$ and in a kink-breather metastable state
for larger $r_0$.
In the latter, the kink sheds at each "return" a
large part of its energy into bursts which are breather solutions.
We have characterized these waves in terms of their energy, frequency and
velocity. In three and higher dimensions and small $r_0$ 
we observe a kink pulson 
bound state. The three states exist in the $(d,r_0)$ parameter space.
This study shows that radial perturbation opens a channel 
between the kink solutions and the breather solutions. This is particularly
interesting because in one dimension these are completely separated.
This additional term provides therefore a mechanism to "destroy" these 
topological defects and extract the energy they contain.

\begin{figure}
\centerline{ \epsfig{file=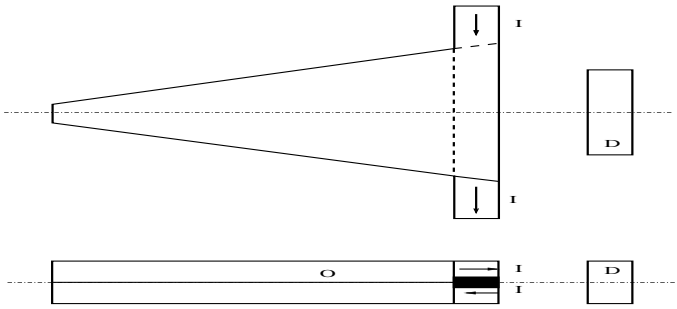,height=4 cm,width=9 cm,angle=0}}
\caption{
Sketch of a microwave generator based on a radial
Josephson window junction. 
}
\label{device}
\end{figure}

In view of applications to 2d Josephson junctions this
could be very useful to generate Terahertz radiation. 
At this time output from these devices is low. Here for small
$r_0$ all the kink energy is converted into radiation. This suggests
the design of a new device based on window Josephson junctions. 
A sketch of the device is shown in Fig. \ref{device}. The top panel
shows a top view of the junction together with the radio-frequency
detector D. Notice the current input similar as in \cite{Carapella02}
with its passive region separating the electrodes.
The bottom panel shows a side view of the system with the oxide layer
O separating the two superconducting films. As the current $I$ is increased
a train of fluxons is formed. These reflect into the narrow end $r=r_0$
and fast breathers are formed which consist in bursts of microwave. Since
all the kink energy is converted into microwaves, we expect this system
to generate much more radiation than a standard flux-flow.

\section{Acknowledgements} 

The authors thank Egor Alfimov, Peter Christiansen and Yuri Gaididei for very
helpful discussions.
J. G. C. thanks the Institute of Mathematical Modeling and the 
Department of Mathematics at the Technical
University of Denmark for their hospitality during several visits.
The authors acknowledge the Centre de Ressources Informatiques de Haute
Normandie where most of the calculations were done.
J. G. C. is on leave from Laboratoire de Math\'ematiques, INSA de 
Rouen, France.

\end{document}